\documentclass[12pt]{article}
\usepackage[centertags]{amsmath}
\usepackage{amsfonts}
\usepackage{amssymb}
\usepackage{amsthm}
\newcommand{\beq}{\begin{equation}}
\newcommand{\eeq}[1]{\label{#1}\end{equation}}
\newcommand{\bea}{\begin{eqnarray}}
\newcommand{\eea}[1]{\label{#1}\end{eqnarray}}

\renewcommand{\Re}{{\rm Re}\,}
\renewcommand{\Im}{{\rm  Im}\,}
\begin{document}
\setlength{\topmargin}{-1cm} \setlength{\oddsidemargin}{0cm}
\setlength{\evensidemargin}{0cm}
\begin{titlepage}
\begin{center}
{\Large \bf Ghosts of Critical Gravity}

\vspace{20pt}

{\large Massimo Porrati and Matthew M. Roberts}

\vspace{12pt}

Center for Cosmology and Particle Physics\\
Department of Physics\\ New York University\\
4 Washington Place\\ New York, NY 10003, USA

\end{center}
\vspace{20pt}

\begin{abstract}
Recently proposed ``critical'' higher-derivative gravities in $AdS_D$ $D>3$ are 
expected to carry logarithmic representation of the Anti de Sitter isometry 
group. In this note, we quantize linear fluctuations of these critical 
gravities, which are known to be either identical with linear fluctuations of 
Einstein's gravity or else satisfy logarithmic boundary conditions at spacial 
infinity. We identify the scalar product uniquely defined by the symplectic 
structure implied by the classical action, and  show that it does not posses 
null vectors. Instead, we show that the scalar product between any two Einstein
modes vanishes, while the scalar product of an Einstein mode with a logarithmic
mode is generically nonzero.  This is the  basic property of logarithmic representation that 
makes them neither unitary nor unitarizable. 
\end{abstract}

\end{titlepage}

\newpage

\section{Introduction}

It has been known for many years that power-counting renormalizable theories
of gravity can be obtained by adding to the Einstein-Hilbert action appropriate
terms, quadratic in the Ricci and Weyl tensors. In the absence of a 
cosmological constant, these theories admit Minkowski space as background, but 
they are also perturbatively non-unitary~\cite{s}. Since the whole point of
having a power-counting renormalizable theory of gravity is to make 
perturbative calculations possible, these theories have been justly abandoned 
long since. Recently, quadratic-curvature actions {\em with} cosmological 
constant were re-examined in four~\cite{lp} and $D$~\cite{d&al} dimensions.
In either case, it was found that there exist a choice of parameters for which
these theories possess one $AdS$ background on which neither massive fields,
nor massless scalars or vectors propagate. Moreover, on the $AdS$ background,
the standard graviton, i.e. the 
massless tensor mode of Einstein-Hilbert gravity, also propagates and 
has vanishing energy (the energy of course depends on the action, not just on 
the form of the mode)~\cite{lp,d&al}. 

Besides those  that satisfies the homogenous Einstein equations on $AdS_D$, 
other tensor modes propagate in the 
``critical'' theory~\cite{af,g&al,b&al}. Their asymptotic behavior at 
space-like infinity differs from standard Einstein-Hilbert modes by 
terms logarithmic in the AdS radial coordinate. A complete set of propagating modes for critical $D$-dimensional gravity was
presented in~\cite{b&al}. 

In this note, we show that there exists an unambiguous
manner to define the energy and the norm of all modes of log gravity.
With that definition, the scalar product of two modes that solve the 
homogenous Einstein equation vanishes. 

Next we come to out main result: the scalar product of a homogenous mode with
some of the logarithmic modes is nonzero. In other words, homogenous modes 
{\em are not} null vectors and cannot be factored out to yield a 
(positive-norm) Hilbert space, except if we restrict the physical space to homogenous modes only, and then factor them out. This procedure leave a profoundly uninteresting theory made only the vacuum state. This picture should be compared to the case of $D=3$, where CFTs possess two copies of the Virasoro algebra. There, in Topologically Massive Gravity (TMG) at the critical point~\cite{lss}, restriction to homogenous modes --which can be promoted to a bona fide non-perturbative constraint on the Hilbert space-- selects the vacuum of one such
algebra,  but allows for nontrivial states of the other~\cite{lss,mss}.
  
This result shows that the ``critical'' (a.k.a. log) theory is neither unitary nor does it
contain a unitary subspace other than the vacuum~\footnote{The ghost pole in one-particle exchange amplitudes between physical sources may cancel~\cite{sgt}; this is not enough to rescue unitarity as we shall discuss in the last section.}.  Though the lack of unitarity 
proven here is bad news for log gravity to give a viable quantum theory of 
gravity in $AdS_D$, it is consistent with (and indeed required by) it being dual
to a logarithmic conformal field theory in $D-1$ dimensions. Such a theory 
could be of interest in statistical mechanics; its basic properties are 
summarized in the next section.

\section*{Kinematics of Logarithmic CFTs}

The isometry group of $AdS_D$ is $SO(2,D-1)$, which is also the conformal
group in $D-1$ space-time dimensions. So, the Hilbert space~\footnote{By 
Hilbert space we mean here a vector space $H$, 
endowed with a non-degenerate but not necessarily positive bilinear form 
$\langle u,v \rangle $ such that $u\in H,\, v\in H \mapsto \langle u,v \rangle\in \mathbb{C}$.} of a consistent 
quantum gravity in $AdS_D$ decomposes into a direct sum of representations of
$SO(2,D-1)$. 

If we do not demand that the representation be unitary, then the Hilbert 
space, $H$, can contain a reducible but {\em indecomposable} representation. 
Let us consider in detail
the case of $AdS_4$. Its isometry group, $SO(2,3)$, admits a Cartan 
decomposition into four positive roots $E^{\alpha_a}$, $a=1,2,3,4$, four 
negative roots $E^{-\alpha_a}$, and two Cartan generators $H_1$, 
$H_2$.~\footnote{The basis used in ref.~\cite{b&al} is: $\alpha_1=(-1,1)$,
$\alpha_2=(0,1)$, $\alpha_3=(-1,-1)$, $\alpha_4=(-1,0)$.}.
In the Cartan basis the $SO(2,3)$ algebra is:
\beq
[H_i,H_j]=0, \qquad [H_i,E^{\alpha_a}]=\alpha_a^i E^{\alpha_a}, \qquad 
[E^{-\alpha_a},E^{\alpha_a}]={2\over |\alpha_a|^2 } \alpha_a \cdot H ,
\eeq{m1}

with $|\alpha_a|^2=\sum_{i=1}^2 (\alpha_a^i)^2$. 

$SO(2,3)$ representations with energy $H_1$ bounded below posses a ground state $\psi$, annihilated by all
$E^{\alpha _a}\psi=0$. Lowest weight vectors can also define logarithmic representations if they are not eigenstates
of $H_1$, but obey instead~\footnote{One could consider in principle also poly-logarithmic representations defined by
$(H_1-E_0)\psi^k=\psi^{k+1}$, $k=0,...,n$.}
\beq
H_1\psi = E_0 \psi + \phi, \qquad H_1\phi=E_0\phi.
\eeq{m2}
When the generators $H_i$, $E^\alpha$, are self-adjoint with respect to a scalar product, $\langle \; , \; \rangle$, not necessarily positive definite, then the vector $\phi$ has zero norm:
\beq
0=\langle\psi , (H_1-E_0)^2 \psi \rangle= \langle (H_1-E_0)\psi , (H_1 -E_0) \psi \rangle = \langle\phi , \phi \rangle.
\eeq{m3}
The standard procedure to obtain a non-degenerate scalar product is to identify vectors modulo null vectors. So, if the vector $\phi$ is null, i.e. if $\langle \chi, \phi\rangle=0$ for all $\chi$ in the representation $V$,  eqs.~(\ref{m2}) actually define a
standard lowest-weight representation on the quotient space. So, to obtain a truly new representation, the scalar product of $\phi$ with some vector $\chi \in V$ (such that $\langle\chi,\chi\rangle\neq 0$) {\em must} be non-vanishing. 

Of course, the latter property is incompatible with unitariy, i.e. with the scalar product being positive definite. For if 
$\langle \phi, \chi \rangle =A\neq 0$, the norm of $z \phi + \chi$ is $\langle\chi,\chi\rangle + 2\Re A z$, which can have either sign when $z$ ranges over the complex plane. 

Two dimensional logarithmic conformal field theories too are characterized by having zero norm non-null vectors~\cite{g} (for a review see \cite{Flohr:2001zs}).  Topologically Massive Gravity (TMG) at the critical point, which is conjectured to be dual to a logarithmic CFT,  indeed contains such vectors~\cite{stvr,gs,gj}. 

Four dimensional critical gravity is in many ways the higher dimensional analog of TMG at the critical point; so, the next question to ask is: does critical gravity too contain logarithmic representations of $SO(2,3)$?
\section*{The Example of $D=4$ Critical Gravity}

Ref.~\cite{b&al} gives a complete set of modes for critical gravity. With some obvious changes of notations and simplifications, the $4D$ action of~\cite{b&al} is
\beq
S= {1\over 16\pi G} \int d^4x \sqrt{-g} \left [ R -2\Lambda - {1\over 2}f^{\mu\nu}G_{\mu\nu} +{m^2 \over 8}(f^{\mu\nu}f_{\mu\nu} -f^2)\right],
\eeq{m4}
with $G_{\mu\nu}$ the Einstein tensor and $f_{\mu\nu}$ an auxiliary symmetric tensor field. Elimination of $f_{\mu\nu}$ through its algebraic equations of motion gives an action quadratic in curvatures. Critical gravity is obtained when the cosmological constant is
\beq
\Lambda =-3m^2.
\eeq{m5}
In $4D$, $\Lambda$ is the usual cosmological constant and action~(\ref{m4}) admits an Anti de Sitter background $\bar{g}_{\mu\nu}$ with
$R^{\mu\nu}_{\,\,\rho\sigma}=(\Lambda/3)(\delta^\mu_\rho \delta^\nu_\sigma - \mu \leftrightarrow \nu) $. 

Expanding $g_{\mu\nu}$ and $f_{\mu\nu}$ around the AdS background as 
\beq
g_{\mu\nu}=\bar{g}_{\mu\nu} + h_{\mu\nu},\qquad f_{\mu\nu} =-2(\bar{g}_{\mu\nu} + h_{\mu\nu}) + {2\over 3m^2}k_{\mu\nu},
\eeq{m6}
Action~(\ref{m4}) reduces to a constant term plus the quadratic action~\cite{b&al}
\beq
6m^2 S_2=\int d^4 x \sqrt{-\bar{g}}\left[ 2h^{\mu\nu} {\cal G}_{\mu\nu}(k) -{1\over 3} (k^{\mu\nu}k_{\mu\nu} -k^2)\right].
\eeq{m7}
The linearized Einstein operator $ {\cal G}_{\mu\nu}$ reduces to $-(1/2) (\Box +2 m^2)$ on transverse-traceless modes.
The equations of motion following from action~(\ref{m7}) are
\beq
{\cal G}_{\mu\nu} (h) ={1\over 3}(k_{\mu\nu} - \bar{g}_{\mu\nu} k), \qquad {\cal G}_{\mu\nu} (k)=0.
\eeq{m7a}
Thanks to the Bianchi identity, $k_{\mu\nu}$ is transverse and traceless.
In the gauge $D^\mu h_{\mu\nu} -D_\nu h =0$ eqs.~(\ref{m7a}) become
\beq
-{1\over 2}(\Box +2 m^2)h_{\mu\nu}= {1\over 3} k_{\mu\nu}, \qquad -{1\over 2}(\Box +2 m^2)k_{\mu\nu}=0.
\eeq{m7b}

In global coordinates (co-latitude $\theta$, longitude $\phi$, radius $\rho$, and time $t$) the $AdS _4$ metric is 
\beq
m^2 ds^2= -\cosh^2(\rho) dt^2 + d\rho^2 + \sinh^2(\rho)[d\theta^2 + \sin^2(\theta) d\phi^2].
\eeq{m8}
The Cartan generators of $SO(2,3)$ are $H_1=i\partial_t$, $H_2=-i\partial_\phi$~\cite{b&al}. The explicit expressions for the $E^\alpha$ generators are also given in~\cite{b&al}. Imposing $E^{\alpha_a}\psi=0$ one finds two particularly interesting classes of solutions to~(\ref{m7b}). 

One, $\psi^E_{\mu\nu}$, solves the homogeneous linearized Einstein equation ${\cal G}_{\mu\nu}(\psi^E)=0$. In global coordinates it has the form $\psi^E_{\mu\nu}=e^{-iE_0t + 2i\phi}F(\rho,\theta)_{\mu\nu}$. For large $\rho$ one finds
\beq
 F(\rho,\theta)_{\rho \rho}\sim e^{-(E_0+2)\rho}, \qquad F(\rho,\theta)_{\rho *}\sim e^{-E_0\rho}, \qquad 
F(\rho,\theta)_{**}\sim e^{-(E_0-2)\rho}.
\eeq{m9}
Here $*$ denotes coordinates other than $\rho$ and we did not spell out the $\theta$ dependence in $F$.
Normalizability of $\psi^E$ for $\rho\rightarrow\infty$ gives $E_0=3$~\cite{b&al}.

The other one is 
\beq
f(t,\rho) \psi^E_{\mu\nu}, \qquad f(t,\rho)= i t +\log \sinh \rho .
\eeq{m8a}
It obeys eq.~(\ref{m7a}) with $k_{\mu\nu}=-(9m^2/2)\psi^E_{\mu\nu}$~\footnote{Our inhomogeneous mode equals
that in~\cite{b&al} plus a homogenous mode.} .

After this brief review of the result of~\cite{b&al} we come to the definition of the scalar product and energy for linearized critical gravity. 

\section*{The Inner Product of Quadratic Theories}

Consider a general quadratic action 
\beq
S= \int dt \frac{1}{2} ( -\dot{q}^TL\dot{q} + + \dot{q}^TQq +q^TK q ),
\eeq{m10}
where $L$ and $K$  are symmetric matrices while $Q$ is antisymmetric and commuting with $L$: $[L,Q]=0$. They are defined in terms of a matrix $\Omega$ obeying 
\beq
L\Omega -\Omega^TL=2iQ, \qquad \Omega^TL\Omega =K.
\eeq{m10a}
Reality of $K$ follows from $[L,Q]=0$; together with the first of the equations above, it implies that $L\Omega = S+iQ$, with $S$ a
real symmetric matrix. If $\Omega$ satisfies eqs.~(\ref{m10a}) so does $-\Omega^*$. When some of the eigenvalues of the matrix $\Omega$ coincide, $\Omega$ may be non-diagonalizable but it can always be put it in Jordan form. We choose $\Omega$ to have ``positive frequency'' by demanding that its eigenvalues are positive definite.

Next we split the vector $q$ into $q=2^{-1/2}(A +A^*)$ with  $i\dot{A}=\Omega A$ ($i\dot{A}^*=-\Omega^* A^*$). Eqs.~(\ref{m10a}) guarantee that $A$ solves the equations of motion $L\ddot{A} -2Q\dot{A} + KA$. The canonical momentum $p$
conjugate to $q$ is $p=L\dot{q} -Qq$. The canonical momenta conjugate to $A$, $A^*$  are
\beq
P=L\dot{A}^*-QA^*=iL\Omega^* A^* -QA^*, \qquad P^*= L\dot{A} -QA=-iL\Omega A - QA.
\eeq{m10b}
By using eqs.~(\ref{m10a},\ref{m10b}) it is straightforward to find that the conserved energy is
\beq
H={1\over 2} (\dot{q}^T L \dot{q} + q^T K q )={1\over 2} (P^T\dot{A} + P^{*\, T} \dot{A}^*)
\eeq{m10c}

In canonical quantization, $A^*$ is replaced by the Hermitian conjugate $A^\dagger$ and the non-zero canonical commutation relations are $[A^I, P_J]=i\delta^I_J $, $[A^{\dagger\, I} , P^\dagger_J]=i\delta^I_J$.
The standard Fock vacuum  obeys $A|0\rangle=0$, so a classical positive-frequency solution of the equations of motion, $i\dot{\Phi}(t)= \Omega \Phi(t)$, defines a one-particle state $|\Phi\rangle=-i\Phi^IP_I|0\rangle$, which obeys $A(t)|\Phi\rangle = \Phi(t)|0\rangle$. The
scalar product of two states $|\Phi\rangle$, $|\Psi\rangle$ is then
\beq
\langle \Psi | \Phi \rangle = i(\Psi^{*\,T} L \dot{\Phi} - \Psi^{*\,T}Q\Phi).
\eeq{m11}

An example directly related to action~(\ref{m7}) is 
\beq
L=\left(\begin{array}{cc} 0 & 1 \\ 1 & 0 \end{array} \right), \qquad 
\Omega = \left(\begin{array}{cc} \omega & 0 \\ 1/2\omega & \omega \end{array} \right),\qquad 
A(t)=\left(\begin{array}{c} 2\omega i \alpha e^{-i\omega t} \\  (\alpha t + \beta)e^{-i\omega t}\end{array} \right)
\eeq{m12}
Notice that, though $A(t)$ contains terms linear in $t$, the scalar product is time-independent. Explicitly, on two solutions 
defined by constants $\alpha,\beta$, $\alpha',\beta'$, the scalar product formula~(\ref{m11}) reduces to 
$\langle \Psi | \Phi \rangle= 2\omega \alpha'^* \alpha + 2\omega^2 i(\beta'^*\alpha -\alpha'^*\beta)$. 
States with $\alpha=\alpha'=0$ have vanishing norm, but these states are not null: the norm of state $\alpha=0,\beta\neq 0$ with an $\alpha'\neq 0$ state does not vanish.

Action~(\ref{m7}) has the form~(\ref{m10}). To see that, we can use the fact that $k_{\mu\nu}$ is transverse-traceless and integrate by part in time~\footnote{Initial time and final time configurations are held fixed when varying the action, so we can always add a total {\em time} derivative to the action without changing the equations of motion. Adding a total divergence of space coordinate, instead, changes the boundary conditions at the AdS boundary, so in general it changes the equations of motion by modifying the boundary behavior of the fields.}
\beq
6m^2 S_2=\int d^4 x \sqrt{-\bar{g}}\left[ \bar{g}^{00}D_0 h^{\mu\nu} D_0 k_{\mu\nu}- h^{\mu\nu}\left(\sum_{i=1}^3 D_iD^i+2m^2\right)k_{\mu\nu}
 -{1\over 3} (k^{\mu\nu}k_{\mu\nu} -k^2)\right], 
\eeq{m13}
Now, a straightforward application of formula~(\ref{m11}) gives the scalar product of any two positive-frequency modes of log gravity as
\beq
\langle \psi | \phi \rangle ={3i\over 4} \int d^3 x \sqrt{-\bar{g}} \bar{g}^{00}[ (\Box +2m^2)\psi^*_{\mu\nu}D_0 \phi^{\mu\nu} + \psi^*_{\mu\nu}D_0 (\Box +2m^2)\phi^{\mu\nu}]
\eeq{m14}

Evidently, the scalar product between two solutions of the homogeneous Einstein equations, $\phi=\phi^E$, $\psi=\psi^E$
vanishes. Also, the potentially dangerous term $\propto t$ in the scalar product of two ``log" modes $\phi^s=f(t,\rho)\phi^E$, $\psi=f(t,\rho)\psi^E$  (see definition in eq.~(\ref{m8a})) vanishes.

Crucially, the homogeneous modes $\psi^E$ are {\em not} null vectors, since their scalar product with the log mode  constructed from $\psi^E$, $\phi^s = f(t,\rho)\psi^E$ is nonzero:
\beq
\langle \psi^E | \phi^s \rangle =-{9im^2\over 4} \int d^3 x \sqrt{-\bar{g}} \bar{g}^{00}\psi^{E\, *}_{\mu\nu} D_0 \psi^{E \, \mu \nu}\neq 0.
\eeq{m15}
Non vanishing of eq.~(\ref{m15}) can be proven by a simple direct calculation or by noticing that~(\ref{m15}) is proportional 
to the norm of the transverse-traceless mode $\psi^E_{\mu\nu}$ in standard Einstein 
gravity~\footnote{See e.g. ref.~\cite{gs} for the analogous calculation in 3D chiral gravity.}.

The Einstein modes do have zero scalar product with special logarithmic modes: 
the spin-1 
``Proca'' modes~\cite{b&al}, which have the form $\psi^s_{\mu\nu}=f \psi^E$, 
$\psi^E=D_{(\mu} A_{\nu)}$. Transversality and tracelessness of 
$\psi^E_{\mu\nu}$ hold when $A_\mu$ obeys the massive spin-1 Proca equation
\beq
D^\nu F_{\nu\mu}=6m^2 A_\mu .
\eeq{m15a}
Vanishing of $\langle \psi^E | f D_{(\mu} A_{\nu)}\rangle$ follows immediately from the fact that this quantity is proportional to the Einstein gravity scalar product $\int d^3 x \sqrt{-\bar{g}} \bar{g}^{00}\psi^{E\, *}_{\mu\nu} D_0 D^\mu A^\nu $, which
vanishes because in Einstein gravity transverse modes are orthogonal to pure gauge modes.

In $3D$, the Proca modes are the only inhomogeneous solutions of eq.~(\ref{m7a}), hence 
the Einstein modes are true null vectors that can be modded out to yield 
a positive-metric Hilbert space. Actually, the normalizable Proca modes decay so rapidly at infinity that they too are null in the norm induced by the NMG action. So, the modding out by these null vectors yields a trivial theory in 3D too~\footnote{This fact became clear in conversations with O. Hohm.}. Of course, one can also define the norm of the Proca fields using the Proca action; this is what makes new massive 
gravity~\cite{nmg} at the critical point potentially nontrivial and consistent, at least at linear order.  
In $D>3$, the Proca modes~\cite{b&al} and the truly transverse-traceless spin-2 
logarithmic modes mix under the action of $SO(D-1)\subset SO(2,D-1)$; therefore, one 
cannot consistently keep the Proca modes, which are transverse to the Einstein modes,
 without also keeping the spin-2 modes, which are not.

\section*{Linearized Energy}

A reasonable way to define the energy of a solution which is asymptotically AdS is to construct the linearized stress tensor of a metric perturbation, and integrate the $tt$ component over a spacial slice, which by the Gauss' law constraint reduces to a boundary term. Including the nonlinear terms schematically in the r.h.s of the equations of motion, we find the equations
\beq
\mathcal{G}(k_{\mu\nu})=\Theta_{\mu\nu}^{(NL)},~\mathcal{G}(h_{\mu\nu})-\frac{1}{3}(k_{\mu\nu}-\bar{g}_{\mu\nu}k)=\Xi_{\mu\nu}^{(NL)}.
\eeq{nonlinears}
Note that thanks to the Bianchi identity $\Theta_{\mu\nu}^{(NL)}$ is automatically covariantly conserved, while $\Xi_{\mu\nu}^{(NL)}$ is not. We therefore can use $\Theta=\mathcal{G}(k)$ to construct conserved charges following the method of  \cite{ad} and replacing the metric perturbation $h_{\mu\nu}$ with $k_{\mu\nu}=3\mathcal{G}_{\mu\nu}(h)-\bar{g}_{\mu\nu}\mathcal{G}^\sigma_\sigma(h)$, which gives that the conserved charge for a killing vector $\xi$ is
\beq
E(\bar{\xi})=\frac{1}{8\pi G}\oint  dS_i\sqrt{-\bar{g}}\left[D_\beta K^{0i\nu\beta}-K^{0j\nu i}D_j \right]\bar{\xi}_\nu ,
\eeq{conscharge}
where 
\beq
K^{\mu\alpha\nu\beta}=\frac{1}{2}\left[ \bar{g}^{\mu\beta}H^{\nu\alpha}+\bar{g}^{\nu\alpha}H^{\mu\beta}-\bar{g}^{\mu\nu}H^{\alpha\beta}-\bar{g}^{\alpha\beta}H^{\mu\nu}\right],~H^{\mu\nu}=k^{\mu\nu}-\frac{1}{2}\bar{g}^{\mu\nu} k.
\eeq{KHdefs}
It is clear from this formulation that any solution that is purely an Einstein mode will have $k_{\mu\nu}=0$ and therefore will have all conserved charges vanish, in agreement with other calculations of the energy in critical gravity. To find a nonzero energy we must turn on a nonzero $k_{\mu\nu}$ mode. At large radius the only static, spherically symmetric solution to (\ref{m7a}) is simple in terms of $k_{\mu\nu}=\tilde{k}_{\mu\nu}+D_{(\mu}\xi_{\nu)}$, the Coulomb tail of a mass in AdS ($\tilde{k}_{\mu\nu}$), along with a vector mode\footnote{This is of course not an honest linear diffeomorphism , as by eq.~(\ref{m7a}) and the diff invariance of $\mathcal{G}$ we see that $k$ does not transform.} ($D_{(\mu}\xi_{\nu)}$). The Coulomb taill of a mass in $AdS_4$ behaves as $\tilde{k}_{tt}=-\bar{g}_{tt}\tilde{k}_{\rho\rho}=\tilde{M}/\sinh\rho$. We must include a vector mode to ensure the consistency of the equation for $h$, because $D^\mu\mathcal{G}_{\mu\nu}=0$ and we must therefore require that $D^{\mu}(k_{\mu\nu}-\bar{g}_{\mu\mu}k)=0$ so that the equation of motion is covariantly conserved. This is ensured with $\xi^r=\tilde{M}/6\sinh^2\rho\cosh\rho$, which gives
\beq
k_{tt}=-k_{\rho\rho}\sinh^2\rho\cosh^2\rho=2k_{\theta\theta}=2k_{\phi\phi}/\sin^2\theta=M/\sinh\rho,
\eeq{kfull}
where $M=2\tilde{M}/3$. It is easy to check that the mass as defined in (\ref{conscharge}) is simply $M/ 2 G$. We also note that this nonzero $k_{\mu\nu}$ sources a logarithmic falloff in $h$,
\begin{equation*}
h_{tt}=\frac{M \log\cosh\rho}{3\sinh\rho}+\frac{M\cosh^2\rho}{3}\left(\pi-4\arctan[\tanh(\rho/2)]-\frac{1+\tanh^2\rho}{\sinh\rho}\right),
\end{equation*}
\beq
h_{\rho\rho}=\frac{M \log\cosh\rho}{3\cosh^2\rho\sinh\rho}.
\eeq{hlog}
We can also turn on a homogeneous mode in $h$, that is one satisfying $\mathcal{G}(h)=0$, but this will clearly not contribute to the energy.

\section*{Miscellaneous Remarks}

The one-particle Hilbert space obtained by quantizing critical gravity on its AdS background splits into a sum of Einstein modes and log
modes: $H=H^s \oplus H^E$. The norm $N$ on such space vanishes on $H^E$ but it is off diagonal. Schematically, for
$|\psi\rangle = |\psi^s \rangle \oplus | \psi^E\rangle$, $|\phi\rangle = |\phi^s \rangle \oplus | \phi^E\rangle$, one has:
\beq
\langle \psi | N |\phi \rangle =\left(\langle\psi^s|, \langle\psi^E| \right) \left( \begin{array}{cc} 1 & \alpha \\ \alpha & 0 \end{array}\right)\left(\begin{array}{c} |\phi^s\rangle \\ |\phi^E\rangle \end{array} \right), \qquad \alpha \neq 0.
\eeq{m16}
A computation of the energy using formula~(\ref{m10c}) shows that, on a mode of frequency $\omega$, its {\em unnormalized} expectation value is proportional to the norm $N$: $\langle \psi | H |\phi \rangle=\omega \langle \psi | N |\phi \rangle$. The Fock space contain only
states with positive frequency, so that whenever the norm is nonzero, the normalized energy, 
$\langle \psi | H |\phi \rangle/ \langle \psi | N |\phi \rangle$, is positive and equal to the
frequency $\omega$, as expected.  When the norm vanishes, by consistency one must define the energy so that it also vanishes. Restricting the multi-particle Hilbert space to the Einstein modes  is thus equivalent to selecting the zero-energy
sub-sector of critical gravity, in perfect analogy with  3D chiral gravity~\cite{mss}. To obtain a properly defined Hilbert space, the restriction to Einstein modes must be followed by modding out by null states.

Unlike 3D chiral gravity, here the end result of this procedure leaves only one state in the theory, the Fock vacuum.

The results described so far were obtained by studying linearized critical gravity. Yet, the general structure we found can be promoted to a full non-linear analysis. In particular, the vanishing of energy in solutions which asymptotically become Einstein modes has been shown to hold for black holes too in~\cite{lp,d&al}. A general proof that all asymptotically-Einstein solutions have zero energy should be possible using the general definition of energy in quadratic-curvature gravities given in~\cite{dt}. Conversely, in the previous section we just proved that non-vanishing energy requires the asymptotic behavior of log modes.

It was conjectured in~\cite{a&al} and proven in~\cite{aeg} that TMG in $3D$ is perturbatively stable around
certain warped $AdS_3$ vacua, even for non-critical values of the Chern-Simons coupling constant. Stability is
achieved by restricting the asymptotic boundary conditions in such a way that only boundary gravitons of one 
chirality propagate. This is in perfect analogy with the behavior of 
TMG at the critical point on non-warped $AdS_3$.
In this paper, we argued that in $D>3$, restriction to the Einstein asymptotics can give us only a trivial 
Hilbert space made only of the 
vacuum. So, if an analog to the results of~\cite{a&al,aeg} could be found for $D>3$ higher-derivative gravity, 
it would probably require 
asymptotics on fields that leave no physical state beyond the vacuum.

It may happen that negative-norm ghost poles cancel against some positive norm poles in one-particle exchange diagrams between physical sources~\cite{sgt}. This is not enough to ensure unitarity, since in the full non-linear theory defined by eq.~(\ref{m4}) ghosts can obviously also be pair-produced. Moreover, even at the level of one-particle diagrams, one must still mod out by null states, arriving again at an empty theory.

Finally, we must point out explicitly that the Hilbert space of Einstein modes is the only physically meaningful subspace without negative norm states (unfortunately, it also has no positive norm states). Obviously, by diagonalizing the metric
$N$ in eq.~(\ref{m16}) one can obtain a positive-norm subspace. Unfortunately, that subspace is not closed under $SO(2,3)$ transformations, since logarithmic
representations are indecomposable. It is amusing to check explicitly this property by verifying that a positive-norm 
subspace is not even closed under time evolution. In fact, its vectors must be linear combinations of log modes and gravity modes. Consider in particular a mode
of frequency $\omega$:
$|\psi\rangle = |\psi^s\rangle + |\phi^E\rangle$, $\psi^s=f(t,\rho)\psi^E$. Under a time translation $t \rightarrow t+ \tau$, it
transforms into $\psi(t)\rightarrow \psi(t+\tau)=\exp(-i\omega \tau)(\psi +i\tau\psi^E)$. Equation~(\ref{m15}) says that  $\langle \psi^s | \psi^E\rangle = c \langle \psi^E | \psi^E \rangle_E $, where $\langle | \rangle_E$  is the scalar product in Einstein gravity and $c$ is a positive constant; so the scalar product 
$\langle \psi(t+\tau) | N |\psi(t+\tau) \rangle$ is independent of $\tau$.  Now consider the linear combination $\psi(t) + C\psi(t+\tau)$. Its norm is
\bea
 \langle \psi(t) + C\psi(t+\tau) | N | \psi(t) + C\psi(t+\tau) \rangle  &=& (CC^* + 1)  \langle \psi | \psi \rangle + 2\langle \psi | \psi \rangle \Re (C  e^{-i\omega \tau} ) + \nonumber \\ &&- 2c\langle \psi^E | \psi^E\rangle_E \Im (\tau  C e^{-i\omega \tau}).
\eea{m17}
The right hand side in this equation becomes negative for $C\neq 0$ and $\tau$ large.
\subsection*{Acknowledgments}
It is a pleasure to thank Olaf Hohm, Chris Pope and Eric Bergshoeff for useful correspondence and discussions. 
M.P. is supported in part by NSF grant PHY-0758032, and
by ERC Advanced Investigator Grant n.226455 {\em Supersymmetry,
Quantum Gravity and Gauge Fields (Superfields)}; M.M.R. is supported by the Simons Postdoctoral Fellowship Program.

\end{document}